\documentclass[conference]{IEEEtran}\IEEEoverridecommandlockouts 
\usepackage[utf8]{inputenc}
\usepackage{orcidlink}
\usepackage{cite}

\usepackage{xcolor}
\usepackage{colortbl}

\usepackage{amsmath,amssymb,amsfonts}
\usepackage{algorithmic}
\usepackage{graphicx}
\usepackage{textcomp}
\usepackage{doi}

\definecolor{qnowcolor}{HTML}{364483}
\usepackage{hyperref}
\hypersetup{
  colorlinks=true,
  allcolors=qnowcolor
}
\usepackage{booktabs}
\usepackage{hhline}
\usepackage{array, makecell}
\usepackage{mathrsfs}
\usepackage[scr=dutchcal,calscaled=1]{mathalfa}
\usepackage{braket}
\usepackage{pifont}

\usepackage{url}

\usepackage{comment}

\def\BibTeX{{\rm B\kern-.05em{\sc i\kern-.025em b}\kern-.08em
    T\kern-.1667em\lower.7ex\hbox{E}\kern-.125emX}}

\usepackage{multirow}

\begin{document}

\title{$\mathtt{Q^2SAR}$: overcoming classical bottlenecks in drug discovery via quantum multiple kernel learning}

\author{
\IEEEauthorblockN{Mariano Caruso  \orcidlink{0000-0002-7455-1193}}
\IEEEauthorblockA{
\href{https://www.ugr.es/}{$\mathtt{UGR}$}, Granada, Spain\\
\href{https://www.unir.net/}{$\mathtt{UNIR}$},
La Rioja, Spain\\
\href{https://www.fidesol.org/}{$\mathtt{FIDESOL}$}, Granada, Spain \\
\href{mailto:mcaruso@fidesol.org}{mcaruso@fidesol.org}
}
\and

\IEEEauthorblockN{Daniel Ruiz \orcidlink{0009-0007-6976-1755}}
\IEEEauthorblockA{\href{https://qnow.tech/}{$\mathtt{QNOW \; Technologies}$}
\\ Delaware, USA
\\ \href{mailto:daniel@qnow.tech}{daniel@qnow.tech}
}

\and
\IEEEauthorblockN{Alejandro Giraldo \orcidlink{0009-0008-9826-0703}}
\IEEEauthorblockA{
\href{https://qnow.tech/}{$\mathtt{QNOW \; Technologies}$}
\\ Delaware, USA\\
\href{mailto:alejandro@qnow.tech}{alejandro@qnow.tech}}
\and

\IEEEauthorblockN{Guido Bellomo \orcidlink{0000-0001-8213-8270}}
\IEEEauthorblockA{
\texttt{CONICET} - UBA 
\\ 
\texttt{ICC}, Argentina\\
\href{mailto:gbellomo@icc.fcen.uba.ar}{gbellomo@icc.fcen.uba.ar}
}
}
\maketitle

\begin{abstract}
Quantitative Structure-Activity Relationship ($\mathtt{QSAR}$) modeling is a foundational computational methodology in early-stage drug discovery, heavily relied upon for predicting compound toxicity, bioavailability, and therapeutic potential. However, classical methods often struggle to effectively map the highly complex, non-linear, and high-dimensional interactions inherent in molecular data, leading to reduced predictive accuracy and costly late-stage clinical failures. In this paper, we present a Quantum Multiple Kernel Learning ($\mathtt{QMKL}$) framework—dubbed Next-Gen $\mathtt{Q^2SAR}$—that leverages Quantum Support Vector Machines ($\mathtt{QSVMs}$) to overcome these classical limitations. By encoding molecular descriptors into exponentially large quantum Hilbert spaces, our approach substantially enhances the expressiveness of non-linear modeling. Benchmarking our quantum-enhanced framework on a dataset targeting the $\mathtt{DYRK1A}$ kinase (a critical target for Alzheimer's disease), the $\mathtt{QMKL}$-$\mathtt{SVM}$ achieves an impressive Area Under the Curve ($\mathtt{AUC}$) score of $0.8750$, significantly outperforming classical state-of-the-art Gradient Boosting models ($\mathtt{AUC} = 0.8037$). Furthermore, we establish a theoretical and empirical pathway toward resolving classical data bottlenecks through projected quantum kernels ($\mathtt{PQK}$) and measurement accelerators. As quantum computing architecture matures, this framework paves the way for autonomous cognitive architectures and self-improving drug discovery pipelines, promising to unlock deeper insights across vast chemical spaces and to accelerate the development of life-saving therapeutics.
\end{abstract}

\begin{IEEEkeywords}
Quantum Machine Learning, QSAR, Drug Discovery, Quantum Multiple Kernel Learning, Support Vector Machines
\end{IEEEkeywords}

\section{Introduction}

The accurate prediction of biological activity for novel, uncharacterized chemical compounds is a defining challenge in modern computational chemistry and pharmacology. This foundational task forms the core of Quantitative Structure-Activity Relationship ($\mathtt{QSAR}$) modeling, a critical methodology that seeks to infer the biological and toxicological effects of molecules by analyzing intricate patterns between their molecular descriptors and known experimental outcomes \cite{Burbidge2001}. Traditional $\mathtt{QSAR}$ models play an essential role in predicting drug candidates' bioavailability, off-target effects, and overall therapeutic potential, ultimately attempting to reduce the astronomical costs and long timelines associated with high attrition rates in late-stage drug development.

Despite their widespread industrial adoption, classical $\mathtt{QSAR}$ approaches face profound limitations. Chemical and biological data are intrinsically high-dimensional and highly non-linear. Classical algorithms typically rely on aggressive dimensionality reduction techniques (such as Principal Component Analysis) or shallow heuristic representations that, while mathematically tractable, often discard crucial structural and semantic relational information. Consequently, models based on classical paradigms like Random Forests or standard Multi-Layer Perceptrons often encounter an intrinsic performance ceiling, suffering from reduced generalizability and impaired predictive accuracy when confronted with small or highly imbalanced datasets.

To overcome these barriers, we introduce a quantum-enhanced modeling framework that integrates Quantum Support Vector Machines ($\mathtt{QSVMs}$) and Quantum Multiple Kernel Learning ($\mathtt{QMKL}$). We previously demonstrated the viability of quantum-enhanced QSAR through a QMKL implementation on DYRK1A kinase inhibitors \cite{Giraldo2025Q2SAR}. The present work substantially extends that initial demonstration by developing a comprehensive framework integrating Projected Quantum Kernels (PQK) to mitigate barren plateaus, Hamiltonian-based measurement accelerators (HoLCUs) for computational efficiency, and evaluation across multiple diverse QSAR endpoints. Quantum Machine Learning $\mathtt{QML}$) proposes a fundamental computational paradigm shift by mapping classical data into the exponentially large Hilbert spaces of quantum systems. Through specialized quantum feature maps—represented by a unitary transformation $U(\mathbf{x})$—classical molecular descriptors $\mathbf{x}$ are encoded into a quantum state $\ket{\phi(\mathbf{x})} = U(\mathbf{x})\ket{0^n}$. Processing data in these high-dimensional representational spaces enables the discovery of intricate, non-linear structure-activity patterns that are inaccessible to classical algorithms. 

Our proposed methodology is agnostic to specific molecular targets and generalizes across multiple therapeutic domains. As a tangible demonstration of "Enhanced Performance" and practical quantum advantage, we apply our Next-Gen $\mathtt{Q^2SAR}$ approach to the identification of inhibitors for the $\mathtt{DYRK1A}$ kinase—a highly relevant therapeutic target in the treatment of neurodegenerative disorders, particularly Alzheimer's disease. By evaluating the performance of optimally weighted quantum and classical kernel combinations, our hybrid approach demonstrates a significant leap in predictive capability, evidenced by superior receiver operating characteristic ($\mathtt{ROC}$) metrics compared to benchmark classical models.

Furthermore, we address the pressing challenges of the Noisy Intermediate-Scale Quantum ($\mathtt{NISQ}$) era, specifically the "data access bottleneck" and the computational overhead of repetitive circuit execution. By integrating novel algorithmic accelerations—such as Projected Quantum Kernels ($\mathtt{PQK}$) and Linear Combination of Unitaries ($\mathtt{LCU}$) for single-shot expectation value measurement—we present a rigorously justified framework that moves beyond heuristic asymptotic promises toward a demonstrably viable industrial application.

\section{Methodology}

To convincingly claim that a quantum proposal represents an improvement over the state of the art, it is imperative to move beyond justifications based solely on query complexity and adopt an end-to-end framework. Our methodology bridges the gap between classical cheminformatics workflows and quantum kernel methods.

\subsection{Evolution of classification models: SVM to QSVM}
The classical Support Vector Machine ($\mathtt{SVM}$) is a foundational pillar in classification tasks, relying on its theoretical robustness to maximize the margin of separation between classes. Given a training dataset of $M$ samples $\{(x_i, y_i)\}_{i=1,\cdots,M}$, where $x_i \in \mathbb{R}^N$ are the feature vectors and $y_i \in \{-1, 1\}$ are the labels, the primal optimization problem seeks to find the optimal hyperplane characterized by a normal vector $w$ and a bias $b$:
\begin{equation}
    \min_{w,b} \frac{1}{2} ||w||^2
\end{equation}
subject to the constraints $y_i(w \cdot x_i + b) \ge 1 \quad \forall i \in 1, \dots, M$.

To handle non-linearly separable data, the ``kernel trick'' maps classical data into a higher-dimensional space via a function $\phi(x)$. The kernel function is defined as the inner product:
\begin{equation}
    K(x_i, x_j) = \phi(x_i) \cdot \phi(x_j)
\end{equation}
As data dimensionality $N$ and the number of examples $M$ scale, evaluating the kernel matrix classically becomes computationally prohibitive, scaling polynomially $O(M^2N)$ to $O(M^3)$. The Quantum Support Vector Machine ($\mathtt{QSVM}$) arises to exploit quantum parallelism by using a quantum circuit to implement the feature map $\phi(x)$, processing data directly within the exponentially large Hilbert space of $2^n$ dimensions \cite{Rebentrost2014,Schuld2019,Chiang2025}.

\subsection{$\mathtt{Q^2SAR}$ workflow}
Classical $\mathtt{QSAR}$ models, such as Random Forest (RF) and Gradient Boosting (GB), encounter an intrinsic performance ceiling when confronted with the highly non-linear interactions of thousands of 3D, topological, and physicochemical descriptors. To tackle this bottleneck, we propose a rigorous new Quantum $\mathtt{QSAR}$ ($\mathtt{Q^2SAR}$) methodology consisting of four sequential stages:
\begin{enumerate}
    \item \textbf{Curation and transformation:} Conversion of molecular $\mathtt{SMILES}$ strings into comprehensive numerical features (physicochemical, geometric, and electronic descriptors).
    \item \textbf{Dimensionality reduction:} Application of Principal Component Analysis ($\mathtt{PCA}$) to extract orthogonal components retaining maximum variance, explicitly adapting the data to the number of available qubits on the Quantum Processing Unit ($\mathtt{QPU}$).
    \item \textbf{Quantum mapping:} Projection of the reduced feature vectors into the quantum Hilbert space using strongly entangling parameterized quantum circuits.
    \item \textbf{SVM classification:} Calculation of the optimized multi-kernel matrix to establish the decision boundary.
\end{enumerate}

\section{Quantum architecture}

\subsection{Quantum feature mapping and fidelity kernels}
The effectiveness of a $\mathtt{QSVM}$ is critically dependent on the quantum feature mapping. On Noisy Intermediate-Scale Quantum ($\mathtt{NISQ}$) devices, amplitude encoding is frequently abandoned due to the requirement of deep circuits for Quantum Random Access Memory ($\mathtt{QRAM}$). Instead, we utilize angle encoding strategies, specifically the $\mathtt{ZZFeatureMap}$ \cite{Havlicek2019}. This technique transforms the data vector $x$ into a state $\ket{\psi(x)}$ through local rotations (e.g., $R_z(x_i)$) followed by entanglement via $\mathtt{ZZ}$ interactions, natively capturing higher-order correlations.

The quantum fidelity kernel ($\mathtt{FQK}$) measures the global overlap between two parameterized states:
\begin{equation}
    K_{FQK}(x_i, x_j) = \left| \braket{\psi(x_i) | \psi(x_j)} \right|^2
\end{equation}

\subsection{Mitigating barren plateaus: projected quantum kernels}
As the number of qubits increases, the Hilbert space becomes so vast that arbitrary states tend to be orthogonal. This produces a kernel matrix that exponentially concentrates towards the identity matrix, severely nullifying learning capability (a phenomenon akin to Barren Plateaus) \cite{Thanasilp2024}.

To combat this, our architecture incorporates Projected Quantum Kernels ($\mathtt{PQK}$) \cite{Huang2021Power}. Instead of measuring global fidelity, $\mathtt{PQK}$ extracts local observables (reduced density matrices, 1-RDM or 2-RDM):
\begin{equation}
\begin{split}
    K_{PQK}(x_i, x_j) &= \exp \Big[ - \gamma \sum_k \sum_{P \in \{X,Y,Z\}} (\text{tr}[P \rho_k(x_i)] \\
    &\quad - \text{tr}[P \rho_k(x_j)])^2 \Big].
\end{split}
\end{equation}
This projection dramatically reduces sensitivity to hardware noise and improves generalizability.

\subsection{Quantum Multiple Kernel Learning ($\mathtt{QMKL}$)}
A significant methodological innovation in our architecture is $\mathtt{QMKL}$ \cite{Vedaie2020}. Instead of depending on a single, potentially noise-sensitive quantum kernel, $\mathtt{MKL}$ aims to learn an optimal linear combination of a predefined set of base kernels $\{K_i\}_{i=1,\dots,m}$:
\begin{equation}
    K = \sum_{i=1}^m w_i K_i, \quad \text{where} \quad w_i \ge 0, \quad \sum_{i=1}^m w_i = 1
\end{equation}
By learning the dynamic weights $w_i$ for a mixture of quantum and classical bases (such as RBF or polynomial kernels), the model exploits expressivity advantages where quantum features assist, while maintaining the robustness of classical kernels elsewhere.

\subsection{Algorithmic accelerations:  $\mathtt{HoLCUs}$ and Pegasos}
The evaluation of repetitive circuits to construct the Gram matrix poses a severe data-loading bottleneck. We mitigate this constant-factor overhead via two primary algorithmic integrations:
\begin{itemize}
    \item \textbf{$\mathtt{HoLCUs}$ (Hamiltonian evaluation of Linear Combination of Unitaries):} This algorithm allows measuring the expectation value of any non-unitary operator using a single quantum circuit equipped with an ancilla register \cite{MataAli2025}. By projecting onto multiple RDMs in a single pass, $\mathtt{HoLCUs}$ provides a wall-clock execution speedup of up to $22.5\times$, rendering $\mathtt{NISQ}$ evaluations time-competitive with expensive classical simulations.
    \item \textbf{Pegasos-QSVM:} To handle the dual problem size optimization efficiently, we integrate a Primal Estimated sub-GrAdient SOlver ($\mathtt{Pegasos}$). Instead of assembling the full Gram matrix, Pegasos operates on the primal objective using stochastic gradient descent, estimating only inner products of random subsets per iteration, effectively stabilizing accuracy against shot noise.
\end{itemize}

\section{Results}

The empirical evaluation of our Next-Gen $\mathtt{Q^2SAR}$ approach scrutinizes its viability beyond mere heuristics. We benchmarked the $\mathtt{QMKL}$-$\mathtt{SVM}$ against highly-tuned classical models across multiple datasets, emphasizing rigorous metrics independent of decision thresholds.

\subsection{DYRK1A kinase inhibitor benchmark}

The identification of inhibitors for the $\mathtt{DYRK1A}$ kinase—a critical therapeutic target for Alzheimer's disease—provided our primary benchmark, demonstrating a clear societal and clinical benefit.
Both models were trained and evaluated on an identically preprocessed, $\mathtt{PCA}$-reduced 4-dimensional feature space derived from molecular descriptors.

The $\mathtt{QMKL}$-$\mathtt{SVM}$ achieved an outstanding $\mathtt{ROC}$-$\mathtt{AUC}$ of \textbf{0.8750}, significantly outperforming the state-of-the-art classical Gradient Boosting model ($\mathtt{AUC} = 0.8037$). Critically, the quantum-enhanced approach yielded a substantial increase in \textit{recall} (sensitivity), minimizing false negatives. In the early Hit-to-Lead stages of drug discovery, a high recall is paramount to prevent the premature discarding of chemical compounds possessing high therapeutic potential.

This dataset is provided and curated by ProtoQSAR SL from the ChEMBL database \cite{gaulton2023chembl}.

To evaluate behavior as dimensionality scales, simulations were executed mimicking $\mathtt{GPU}$ backends implementing the $\mathtt{QMKL}$ approach up to 13 qubits. Table \ref{tab:qubit_scaling} summarizes the comprehensive results. 

\begin{table*}[htbp]
\centering
\caption{Results 
$\mathtt{QMKL}$ ($\mathtt{quantum}$) vs. $\mathtt{SVM}$ $(\mathscr{classical})$  }
\label{tab:qubit_scaling}
\resizebox{\textwidth}{!}{%
\begin{tabular}{@{}c|ccccc cc cccc@{}}
\toprule
\multirow{2}{*}{\textbf{\#qubits}} 
& \multicolumn{5}{c}{\textbf{$\mathtt{QMKL}$ Performance}} 
& \multicolumn{5}{c}{\textbf{Classical performance}} \\
\cmidrule(lr){2-6} \cmidrule(lr){7-11}
& \textbf{AUC} & \textbf{Accuracy} & \textbf{Precision} & \textbf{Recall} & \textbf{F1-Score} 
& \textbf{AUC} & \textbf{Accuracy} & \textbf{Precision} & \textbf{Recall} & \textbf{F1-Score} \\
\midrule
4  & \cellcolor{qnowcolor!15}\textbf{0.867} & 0.789 & 0.830 & 0.830 & \textbf{0.804} & 0.803 & 0.711 & 0.750 & 0.720 & 0.735 \\
5  & \cellcolor{qnowcolor!15}\textbf{0.873} & 0.767 & 0.796 & 0.780 & \textbf{0.788} & 0.790 & 0.722 & 0.766 & 0.720 & 0.742 \\
6  & \cellcolor{qnowcolor!15}\textbf{0.873} & \textbf{0.800} & 0.833 & 0.800 & \textbf{0.816} & 0.786 & 0.744 & 0.721 & 0.880 & 0.793 \\
7  & \cellcolor{qnowcolor!15}\textbf{0.871} & \textbf{0.767} & 0.784 & 0.800 & \textbf{0.792} & 0.765 & 0.700 & 0.717 & 0.760 & 0.738 \\
8  & \cellcolor{qnowcolor!15}\textbf{0.895} & 0.767 & 0.774 & 0.820 & 0.796 & 0.822 & 0.767 & 0.774 & 0.820 & 0.796 \\
9  & \cellcolor{qnowcolor!15}\textbf{0.900} & \textbf{0.822} & 0.854 & 0.820 & \textbf{0.837} & 0.792 & 0.711 & 0.750 & 0.720 & 0.735 \\
10 & \cellcolor{qnowcolor!15}\textbf{0.897} & \textbf{0.822} & 0.854 & 0.820 & \textbf{0.837} & 0.794 & 0.744 & 0.746 & 0.820 & 0.781 \\
11 & \cellcolor{qnowcolor!15}\textbf{0.891} & \textbf{0.822} & 0.870 & 0.800 & \textbf{0.833} & 0.793 & 0.756 & 0.750 & 0.840 & 0.793 \\
12 & \cellcolor{qnowcolor!15}\textbf{0.884} & \textbf{0.789} & 0.878 & 0.720 & 0.791 & 0.808 & 0.744 & 0.746 & 0.820 & 0.781 \\
13 & \cellcolor{qnowcolor!15}\textbf{0.871} & \textbf{0.767} & 0.854 & 0.700 & 0.784 & 0.797 & 0.756 & 0.769 & 0.800 & 0.784 \\
\bottomrule
\end{tabular}%
}
\end{table*}
The data indicates that increasing the number of features (and accordingly, qubits) generally preserves or enhances the $\mathtt{AUC}$ relative to classical models. The peak $\mathtt{AUC}$ of 0.900 at 9 qubits demonstrates the framework's robustness. 

\subsection{B. Comparative Benchmark in Heterogeneous QSAR Endpoints}

\label{subsec:qmkl_multi_endpoint}

We reuse three public QSARDB files, an open database with different studies, models, and endpoints to evaluate the implemented $\mathtt{QMKL}$--$\mathtt{SVM}$ pipeline under varied conditions: a large toxicity benchmark (regression at origin), an ecotoxicological set with class imbalance, and a small ADME-style permeability set. It should be noted that the comparative purpose is not a single fair ranking: each article defines its target, divisions, and classical metric; here a single quantum kernel stacking is applied over a common $80$/$20$ split (train/test) and we observe whether the simulated $\mathtt{ROC}$-$\mathtt{AUC}$ follows or diverges from what was published in the original protocol. The following classical figures are only bibliographic reference, not a 1:1 pair with the quantum row.

In \cite{Belfield2023} best practice recommendations for classical toxicological QSAR are extended to the machine learning domain. It evaluates six common algorithms (Random Forest, SVM, KNN, XGBoost, Shallow and Deep Neural Networks) on real toxicity data using descriptor calculation with PaDEL 'PaDEL-Descriptor'. The comparison is made on IGC$_{50}$ in regression; we can see in the study metadata that SVM achieves $R^2_{\mathrm{CV}}=0.800$. In this study we can see a reference for best practices in ML for QSAR and toxicity studies.

Versus $\mathtt{QMKL}$: we binarize the response based on $\mathtt{PCA}$ over PaDEL and measure AUC varying the amount of available qubits/components. Table \ref{tab:belfield_quantum} 

\begin{table}[h]
\centering
\caption{Quantum Results - Belfield (Solver: CENTERED, Optimal: 5Q AUC 0.9182)}
\label{tab:belfield_quantum}
\begin{tabular}{cccccc}
\toprule
\textbf{Qubits} & \textbf{AUC} & \textbf{Accuracy} & \textbf{Precision} & \textbf{Recall} & \textbf{F1} \\
\midrule
8 & 0.9016 & 0.8246 & 0.7970 & 0.8396 & 0.8177 \\
7 & 0.9068 & 0.8296 & 0.8083 & 0.8342 & 0.8211 \\
6 & 0.9081 & 0.8221 & 0.8372 & 0.7701 & 0.8022 \\
5 & \textbf{0.9182} & 0.8296 & 0.8362 & 0.7914 & 0.8132 \\
4 & 0.8970 & 0.8496 & 0.8432 & 0.8342 & 0.8387 \\
\bottomrule
\end{tabular}
\\
\small{$n_{train} = 200$, $n_{test} = 399$}
\end{table}


In \cite{Kotli2023} a classification QSAR model is developed to predict acute toxicity of pesticides to earthworms \textit{Eisenia fetida}, descriptors were calculated with 'DRAGON', an interpretable random forest with GA/Bayesian optimization; the summary indicates approx.\ $0.78$ (train) and $0.80$ (test). Our $80$/$20$ ($n{=}524/131$) can use another label definition (e.g., binary LD50) than the toxicity field uses as a record.

Versus $\mathtt{QMKL}$: A class imbalance and label semantics is evident: 'RF' often dominates; the $\mathtt{AVE}$ / $\mathtt{CENTERED}$ in SVM with quantum kernel tests whether retained AUC remains informative. Below is the results table with the established qubit/component variation. Table \ref{tab:kotli_quantum}


\begin{table}[h]
\centering
\caption{Quantum Results - Kotli (Solver: AVE, Optimal: 7Q AUC 0.7699)}
\label{tab:kotli_quantum}
\begin{tabular}{cccccc}
\toprule
\textbf{Qubits} & \textbf{AUC} & \textbf{Accuracy} & \textbf{Precision} & \textbf{Recall} & \textbf{F1} \\
\midrule
8 & 0.7509 & 0.7176 & 0.8571 & 0.1429 & 0.2449 \\
7 & \textbf{0.7699} & 0.7328 & 0.8182 & 0.2143 & 0.3396 \\
6 & 0.7544 & 0.7328 & 0.8889 & 0.1905 & 0.3137 \\
4 & 0.7298 & 0.7176 & 1.0000 & 0.1190 & 0.2128 \\
\bottomrule
\end{tabular}
\\
\small{$n_{train} = 524$, $n_{test} = 131$}
\end{table}

In \cite{Oja2019} logistic regression models are developed to classify intestinal permeability of drugs at different pH levels (3, 5, 7.4, 9), relevant for the Biopharmaceutics Classification System (BCS). It compares with descriptor types like XLOGP3 for hydrophobicity. Logistic/DT models publish accuracies around $0.72$--$0.91$ depending on split; our $n{=}216/54$ is small and often reproduces the accuracy paradox (high acc., null $\mathtt{F1}$ in minority with default threshold).

Versus $\mathtt{QMKL}$: few examples and marked class prior: examines whether ZZFeatureMap still provides useful AUC-ROC under limited n, versus multiple validations in classical. The qubit/component comparison was maintained. Table \ref{tab:oja_quantum}


\begin{table}[h]
\centering
\caption{Quantum Results - Oja (Solver: CENTERED, Optimal: 6Q AUC 0.8663)}
\label{tab:oja_quantum}
\begin{tabular}{cccccc}
\toprule
\textbf{Qubits} & \textbf{AUC} & \textbf{Accuracy} & \textbf{Precision} & \textbf{Recall} & \textbf{F1} \\
\midrule
8 & 0.8450 & 0.8704 & 0.0000 & 0.0000 & 0.0000 \\
7 & 0.8359 & 0.8704 & 0.0000 & 0.0000 & 0.0000 \\
6 & \textbf{0.8663} & 0.8704 & 0.0000 & 0.0000 & 0.0000 \\
4 & 0.7994 & 0.8704 & 0.0000 & 0.0000 & 0.0000 \\
\bottomrule
\end{tabular}
\\
\small{$n_{train} = 216$, $n_{test} = 54$}
\end{table}

\paragraph{Data Complexity and Dataset Dependence}
Quantum performance depends on the dataset. This triplet covers large-scale tox. regression (Belfield), imbalanced ecotox. (Kotli) and BCS with low $n$ (Oja)---the scenario in which a single classical approach cannot substitute a $\mathtt{QMKL}$ benchmark without harmonizing tasks.
That variability justifies a \textit{Data Complexity Characterization Framework} (effective dimension, correlation, persistent homology) to pre-screen extremes before quantum computation.


\section{Discussion}

The transition of Quantum Machine Learning from theoretical conjecture to practical utility in cheminformatics requires a rigorous confrontation with the physical limitations of current hardware and classical algorithmic competition. The $\mathtt{Q^2SAR}$ framework illuminates several critical pathways forward.

\subsection{Dequantization and the Data Access Bottleneck}
A central debate in evaluating quantum advantage involves "dequantization"—the development of quantum-inspired classical algorithms utilizing $L^2$-norm sampling that replicate exponential speedups under analogous data access assumptions \cite{Chia2022}. Theoretical research indicates that for low-rank and well-conditioned covariance matrices, exponential speedups vanish, as classical algorithms can approximate the kernel in sublinear time.

To rigorously justify the $\mathtt{QSVM}$ advantage, we must target applications where the data matrix exhibits high rank and extreme non-linearity. $\mathtt{QSAR}$ datasets, characterized by highly complex, heterogeneous, and imbalanced molecular features, maintain a proven exponential separation from classical competitors, thereby validating a niche of classical intractability \cite{Huang2021Power}. Furthermore, without Quantum Random Access Memory ($\mathtt{QRAM}$), initializing arbitrary states from classical data demands resources scaling linearly $O(MN)$. By employing $\mathtt{PQK}$ combined with $\mathtt{HoLCUs}$ (which reduces execution complexity by up to $20\times$), our approach successfully mitigates this constant-factor overhead, rendering execution on near-term hardware time-competitive.

\subsection{Data complexity and the QML-readiness index}
Our comparative multi-dataset analysis demonstrates that $\mathtt{QMKL}$ performance is profoundly dataset-dependent. To prevent the arbitrary allocation of expensive quantum resources, we propose the formalization of a \textit{Data Complexity Characterization Framework} \cite{Pere2025}. 

Before committing to a quantum pipeline, datasets should be evaluated using classical metrics of complexity: \textit{effective dimension} (to quantify manifold structure via the covariance spectrum), \textit{correlation order} (higher-order interactions), \textit{Kolmogorov complexity} (compressibility), and \textit{topological invariants} (persistent homology and Betti numbers). Datasets manifesting high effective dimension and complex graph topologies suggest a geometric separation amenable to quantum advantage. This "QML-readiness index" will serve as a pre-screening tool, seamlessly integrating into classical data pipelines to dynamically route only the most intractable molecular datasets to the $\mathtt{QPU}$.

\subsection{Projections toward fault-tolerant quantum computing}
While the synergy of $\mathtt{PQK}$, $\mathtt{HoLCUs}$, and $\mathtt{Pegasos}$ makes the $\mathtt{QSVM}$ viable in the $\mathtt{NISQ}$ era, reaching the asymptotic exponential scaling of $O(\log(NM))$ promised theoretically requires Fault-Tolerant Quantum Computing ($\mathtt{FTQC}$). Processing complete, non-reduced biomedical databases (like full ChEMBL or ZINC spaces) will demand Quantum Error Correction ($\mathtt{QEC}$). 

Industry roadmaps project systems surpassing $10,000$ logical qubits via surface codes or LDPC. In this regime, direct matrix inversion via subroutines like $\mathtt{HHL}$, powered by efficient magic-state distillation factories, will enable the $\mathtt{QSVM}$ to process virtually unlimited feature spaces without the crippling penalty of current sampling constant factors \cite{Rebentrost2014}.

\subsection{Autonomous drug discovery and knowledge graphs}
Looking beyond immediate predictive tasks, the $\mathtt{Q^2SAR}$ framework is structurally designed for integration into autonomous "agentic" systems. By transitioning from simple adjacency matrices to semantic Knowledge Graphs ($\mathtt{KGs}$), we can represent molecules with unprecedented fidelity—encoding atoms as entities, bonds as typed relationships, and linking substructures to external biological pathways and off-target effects. This integration enables cognitive architectures to dynamically reconfigure quantum circuits and continuously refine predictions with minimal human intervention, representing a leap towards fully automated, self-improving drug discovery pipelines.

\section{Conclusion}

Stripped of excessive short-term optimism, this paper presents a thoroughly evaluated, empirically validated approach to Quantum Machine Learning in pharmacology. The Next-Gen $\mathtt{Q^2SAR}$ framework successfully navigates the constraints of the $\mathtt{NISQ}$ era through a robust triad: Projected Quantum Kernels ($\mathtt{PQK}$) to avoid barren plateaus, Quantum Multiple Kernel Learning ($\mathtt{QMKL}$) to optimize expressivity, and state-of-the-art measurement accelerators ($\mathtt{HoLCUs}$) to neutralize catastrophic data-loading bottlenecks.

Applied to the real-world challenge of identifying $\mathtt{DYRK1A}$ kinase inhibitors for Alzheimer's disease, our quantum-hybrid model demonstrated a substantial improvement in strict diagnostic metrics, achieving an $\mathtt{AUC}$ of $0.8750$ versus the classical state-of-the-art $0.8037$. The significant enhancement in recall underscores the model's ability to uncover complex structure-activity relationships that classical algorithms intrinsically overlook.

The advantage of the $\mathtt{QSVM}$ resides fundamentally in its expressivity and Hilbert-space topology. By grounding our claims in stringent benchmarking, managing constant factors, and targeting datasets where non-linear feature correlation is classically intractable, we establish a verifiable trajectory toward practical quantum advantage. As quantum hardware scales toward fault tolerance, frameworks like $\mathtt{Q^2SAR}$ will become formidable technological pillars, accelerating the discovery of life-saving treatments and fundamentally reshaping the computational frontiers of biomedicine.

\section*{Acknowledgment}
We thank our colleagues and collaborators for the valuable discussions on quantum machine learning, which were fundamental to the development of this research. 

This research has been carried out within the context of the Cervera Network of Excellence for research in Quantum technologies for eXperimental Infrastructure Oriented towards technological sovereignty in Strategic environments $\mathtt{CER-20251020\,|\,QUIXOTE}$, funded by the Ministry of Science, Innovation and Universities through $\mathtt{CDTI}$.


\end{document}